\documentclass[a4paper]{PoS}

\title{Partial quenching and chiral symmetry breaking}

\ShortTitle{Partial quenching and chiral symmetry breaking}

\author{\speaker{Michael Creutz}%
\\ %       \thanks{A footnote may follow.}\\
        Brookhaven National Laboratory\\
        E-mail: \email{mike@latticeguy.net}}

\abstract{Partially quenched chiral perturbation theory assumes that
  valence quarks propagating on gauge configurations prepared with sea
  quarks of different masses will form a chiral condensate as the
  valence quark mass goes to zero. I present a counterexample
  involving non-degenerate sea quarks where the valence condensate
  does not form.}

\FullConference{The 32nd International Symposium on Lattice Field Theory\\
        23-28 June, 2014\\
        Columbia University New York, NY}

\begin{document}

%\section{...}

% to comment out parts of a TeX file
\long \def \blockcomment #1\endcomment{}

Partial quenching is a common approximation used to extract
additional information from sets of field configurations obtained in
dynamical lattice gauge simulations.  The idea is to take these
configurations, generated with some set of fixed ``sea'' quark
masses, and then study quark propagators using different ``valence''
quark masses.  From these propagators, one then constructs ``valence''
bound states and studies their properties.  Of course when the sea and
valence masses are equal, this is just the normal lattice prescription
for obtaining hadronic properties.

The conventional assumption is that, as the valence masses go to zero,
a valence quark condensate will form on which one might expect the
valence pion masses to go to zero with the square root of the valence
quark mass.  The point I wish to make is that in some cases this
assumption can fail.  I will show an example, using two non-degenerate
sea quark masses, where the valence pions do not become massless in
the valence chiral limit.

Consider two non-degenerate quark flavors to which I give
the conventional names``$u$'' and ``$d$.''  The standard chiral
symmetry prediction for the dynamical pions is that their mass is
controlled by the average quark mass
\begin{equation}
M_{\pi}^2\sim {m_u+m_d \over 2}+O(m_q^2).
\end{equation}
Now consider fixing the down quark mass to some non-vanishing value,
$m_d\ne 0$, and take the up quark mass to zero.  Then we expect the
pion mass to remain finite
\begin{equation}
M_{\pi}^2\sim {m_d \over 2}+O(m_q^2).
\end{equation}
In particular we expect no singularity in physics for $m_u$ in the
vicinity of zero.  

\begin{figure*}
\begin{centering}
\includegraphics[width=.5\textwidth]{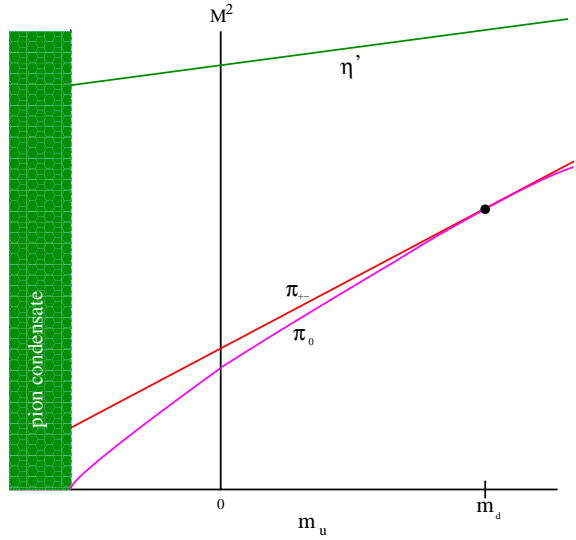}
\caption{
\label{iso2}
A schematic representation of how the pseudo-scalar masses depend on
the up quark mass with a fixed non-vanishing down quark mass.  The
theory maintains a mass gap even at vanishing up quark mass.}
\end{centering}
\end{figure*}

Without a singularity at vanishing up quark mass, it is natural to
imagine continuing the up quark mass to negative values.  In standard
chiral perturbation theory the neutral pion mass continues to drop
until at some point it becomes negative.  At this location a second
order transition is expected into a CP violating phase with an
expectation value for the neutral pion.  This phase was anticipated by
Dashen some time ago \cite{Dashen:1970et}.  For a recent discussions
of these issues see
Refs.~\cite{Creutz:2013hza,Aoki:2014moa,samc,horkel}.  The qualitative
meson mass spectrum as a function of the up quark mass is sketched
qualitatively in Fig.~(\ref{iso2}).

Considerable insight into the nature of the chiral limit in QCD was
provided some time ago by Banks and Casher \cite{Banks:1979yr}.  They
argued that a finite density of small eigenvalues for the Dirac
operator could generate a jump in the condensate $\overline\psi\psi$
as the quark mass passes through zero.  This is exactly as expected
from chiral perturbation theory with degenerate quarks.

The situation with a single massless quark, however, generates a
conundrum for this picture.  As discussed above, as the sea up-quark
mass passes through zero, no singularity is expected.  In particular,
there should be no jump in the up quark condensate $\langle \overline
u u\rangle$ if the down quark mass remains finite.  Following the
Banks Casher argument, the density of up quark eigenvalues
$\rho_u(\lambda)$ must vanish at $\lambda=0$.  For another discussion
of this, see Ref.~\cite{Creutz:2005rd}.

Now we bring in two degenerate valence quarks and consider taking
$m_{val}$ to zero while the dynamical up quark mass is maintained to
vanish.  In the limit that the valence quarks and the up sea-quark
have the same mass, their propagators become identical.
\begin{equation}
D_{val}\rightarrow D_u\qquad \hbox{as}\qquad m_{val}\rightarrow 0.
\end{equation}
We conclude that with a vanishing dynamical up quark
mass we must have
\begin{equation}
\rho_{val}(0) \rightarrow \rho_u(0) = 0.
\end{equation}
Thus the valence quarks can not condense, there is no valence chiral
symmetry breaking, and there is no expectation for the valence pion
mass to go to zero.  This is in direct contradiction to the usual
assumptions of partially quenched chiral perturbation theory.

It is well known that in the fully quenched case, i.e. with no
dynamical quarks, problems arise when the masses for valence quarks
are taken to zero.  In general, configurations will be encountered where
the Dirac operator is not invertible and the propagators do not exist.
In the example presented above, the dynamical quarks suppress such
configurations so the propagators do exist, although their chiral
properties are strongly modified.

While this is the basic result, some technical comments are perhaps in
order.  The discussion above is based on the expectation that the four
dimensional free energy density behaves smoothly as the up quark mass
passes through zero.  The full partition function is the exponential
of the volume times this density, and should be well behaved
throughout the small mass region, including negative mass.  

There are some peculiarities of the path integral at negative quark
mass.  In this situation gauge field configurations can appear for
which the fermion determinant is negative.  This is somewhat
non-intuitive; for instance the topological susceptibility, despite
being the expectation of a square, is itself negative
\cite{Creutz:2013xfa}.  Indeed, the susceptibility diverges to
negative infinity as the Dashen phase is approached.

It should be noted that the expectation
$\langle\overline\psi\psi\rangle$ for the up quark does not vanish at
zero mass.  This does not come from spontaneous chiral symmetry but
rather is a direct consequence of the chiral anomaly.  As discussed in
\cite{Creutz:2005rd} and \cite{Creutz:2006ts} , this expectation
arises from a cancellation of the mass suppression of unit topology
with an inverse mass dependence in this particular observable.  As
such it is tied not to a density of small eigenvalues, but rather to
exact zero modes of the Dirac operator.  This effect is only present
when a single quark becomes massless; with degenerate light quarks
further factors of the mass suppress topology and the Banks-Casher
picture becomes relevant.

At this meeting, Verbaarschot and Wettig \cite{vw} have suggested that
it might be possible for the eigenvalue density at the origin to
remain finite if there is a cancellation at negative mass between
configurations of non-trivial topology.  However, other than the above
contribution to $\langle\overline\psi\psi\rangle$ from unit topology,
effects of higher winding number sectors are suppressed by powers of
the quark mass.  Also, for local observables, at large volume one can
avoid these issues by working in the zero winding number sector
\cite{Aoki:2007ka}.

In summary, I have presented a situation where partially quenched
chiral perturbation theory can fail.  One should be particularly
suspicious of the approach whenever the valence quark masses become
small compared to the average sea quark mass.  This conclusion is a
direct consequence of the anomaly and applies for any valid
lattice fermion formulation.

\end{document}